\documentclass[11pt]{article}
\usepackage{amssymb}
\usepackage[T1]{fontenc}
\usepackage[latin2]{inputenc}
\usepackage[intlimits]{amsmath}
\usepackage{enumerate}
\usepackage[dvips]{graphicx}

\usepackage[all]{xy}

\makeatletter
\linethickness{0.4pt}

\newtheorem{lema}{Proposition}

\numberwithin{equation}{section}

\begin{document}

\title{Second order $q$--difference equations solvable by factorization method}
\author{Alina Dobrogowska and Anatol Odzijewicz\\\\
Institute of Theoretical Physics, University of Białystok\\
Lipowa 41, 15-424 Białystok, Poland\\
{\rm E-mail: alaryzko@alpha.uwb.edu.pl and aodzijew@labfiz.uwb.edu.pl} }

\maketitle

\begin{abstract}
By solving an infinite nonlinear system of $q$-difference equations one constructs a chain of $q$-difference operators. The
eigenproblems for the chain are solved and some applications, including the one related to $q$-Hahn orthogonal polynomials, are discussed.
It is shown that in the limit $q\to1$ the present method corresponds to the one developed by Infeld and Hull.
\end{abstract}

\tableofcontents
\vspace*{1cm}

\section{Introduction}

The discretization of the ordinary differential equations is an important and necessary step toward finding their
numerical solutions. In place of the standard discretization based on the arithmetic progression, one can use  a not less efficient
$q$-discretization related to geometric progression. This alternative method leads to $q$-difference equations, which
in the limit $q\to 1$ correspond to the original differential equations. The theory of $q$-difference equations and the related
$q$-special functions theory have a long history (see e.g. \cite{34}). During the last two decades they have been reviewed because of
the great success of the theory of quantum groups.

The other crucial way of solving ordinary differential equations is based on the factorization method first used by Darboux \cite{9}.
Later the method was rediscovered many times, in particular by the founders of quantum mechanics, see [8] and [11], while studying
the Schr\"odinger equation. We refer to [16] for an exhaustive presentation of the factorization method. In the paper [2], which
is now considered  to be fundamental, Infeld and Hull summarised the quantum mechanical applications of the method. Fixing
an infinite system of Riccati type equations they have constructed a chain of second order differential operators and
proposed some method of solving corresponding eigenproblems.

In this paper we construct the chain \eqref{r30a19} of second order $q$-difference operators by solving an infinite
nonlinear $q$-difference system. This chain depends on a freely chosen function and a finite number of real parameters.
In Section 2 we find a family of eigenvectors for the operators of \eqref{r30a19}. In Section 3 it is shown that $q$-Hahn
orthogonal polynomials, which are $q$-deformation of the classical orthogonal polynomials, form the family
of solutions obtained by our method.
Other examples of solutions obtained by the factorization of $q$-difference equations are presented in Section 4.
Finally passing to the limit $q\to 1$ in \eqref{n1}, \eqref{n2} we obtain some new families of solutions for second order
differential equations. 

\section{Factorized chain of the second order $q$-difference operators}

In this section  we shall consider the sequence of the second order $q$--difference operators
\begin{equation}
\label{a15}
{\bf H}_k= Z_{k}(x)\partial_{q}Q^{-1}\partial_{q}+W_{k}(x)\partial_{q}+V_{k}(x)\; ,
\;\;\;\;\;\;k\in {\mathbb N}\cup \{ 0\}
\end{equation}
acting in the Hilbert spaces ${\cal H}_k$.
By definition ${\cal H}_k$ consist  of the complex valued functions $\psi
:[a,b]_q\rightarrow {\mathbb C}$ defined on the $q$--interval
\begin{equation}
\label{a15a}
[a,b]_q:=\{ q^na:n\in {\mathbb N}\cup \{ 0\}\}\cup \{ q^nb:n\in {\mathbb N}\cup \{ 0\}\}
\end{equation}
and square--integrable, i.e. $\langle \psi | \psi \rangle_k<+\infty$,
with respect to the scalar products
\begin{equation}
\label{a1}
\langle \psi | \varphi \rangle_k:=\int_{[a,b]_q}\overline{\psi (x)}\varphi (x)\varrho_k (x) d_{q}x\;.
\end{equation}
Let us recall (see [34]) that by definition the $q$--derivative is 
\begin{equation}
\partial_q\psi(x)=\frac{\psi(x)-\psi(qx)}{(1-q)x}\;,
\end{equation}
and the $q$--integral on the $q$--interval $[a,b]_q$ is given by
\begin{equation}
\int_{[a,b]_q}\psi(x)d_qx:=\sum_{n=0}^{\infty}(1-q)q^n\left( b\psi(q^nb)-a\psi(q^na)\right)\;.
\end{equation}
If $a=0$ and $b=\infty$ then
\begin{equation}
\int_{0}^{\infty}\psi(x)d_qx:=\lim_{n\rightarrow \infty}\int_{0}^{q^{-n}}\psi(x)d_qx =
\sum_{n=-\infty}^{\infty}(1-q)q^nf(q^n)\;.
\end{equation}
In the case if  $a=-\infty$ and $b=\infty$
\begin{equation}
\int_{-\infty}^{\infty}\psi(x)d_qx:=\lim_{n\rightarrow \infty}\int_{-q^{-n}}^{q^{-n}}\psi(x)d_qx =
\sum_{n=-\infty}^{\infty}(1-q)q^n\left( f(q^n)+f(-q^n)\right)\;.
\end{equation}
In the limit $q\rightarrow 1$ the above definitions correspond to their counterparts 
in standard calculus.

The scalar products (\ref{a1}) are defined by the weight functions  $\varrho_{k}:[a,b]_q\rightarrow {\mathbb R}$, which are
related by the recursion relations
\begin{equation}
\label{a2}
\varrho_{k-1}=\eta_{k}\varrho_{k}\;
\end{equation}
and
\begin{equation}
\label{a3}
\varrho_{k-1}=Q \left( B_{k}\varrho_{k} \right)\;,
\end{equation}
where $\eta_k$, $B_k$ are real valued functions on  $[a,b]_q$ and the operator $Q$ is defined 
by the formula
\begin{equation}
\label{a3a}
Q\varphi (x)=\varphi (qx)\;.
\end{equation}
For the sake of consistency we need to add the conditions
\begin{equation}
\label{a4}
Q\left( B_{k}\varrho_{k} \right) =\eta_{k}\varrho_{k}
\end{equation}
on the functions $\eta_k$ and $B_k$.
Additionally we impose the boundary conditions
\begin{equation}
\label{a5}
B_{k}(a)\varrho_{k}(a)=B_{k}(b)\varrho_{k}(b)=0\;.
\end{equation}
If we introduce the functions
\begin{equation}
\label{a6}
A_{k}(x):=\frac{B_{k}(x)-\eta_{k}(x)}{(1-q)x}\;,
\end{equation}
we can rewrite the formula (\ref{a4}) in the form of a $q$--Pearson equation \cite{4}
\begin{equation}
\label{a7}
\partial_{q} \left( B_{k}\varrho_{k} \right) =A_{k}\varrho_{k}\;.
\end{equation}
In the limit $q\rightarrow 1$, the equation (\ref{a7}) corresponds to the Pearson equation
which is important for the theory of classical orthogonal polynomials \cite{40}.

We say that the operators ${\bf H}_k$ admit a factorization if
\begin{equation}
\label{a18}
{\bf H}_{k}={\bf A}_{k}^{ *}{\bf A}_{k}+a_k
\end{equation}
and
\begin{equation}
\label{3}
{\bf H}_{k}=d_{k+1}^{-1}\left({\bf A}_{k+1}{\bf A}_{k+1}^*+a_{k+1}\right) \;,
\end{equation}
where the annihilation operators ${\bf A}_{k}:{\cal H}_{k}\rightarrow {\cal H}_{k-1}$ are of 
the form
\begin{equation}
\label{a16}
{\bf A}_{k}=\partial_{q} +f_{k}
\end{equation}
and $f_k$ are real valued  functions on the set $[a,b]_q$.
The adjoint operators  ${\bf A}_{k}^{*}: {\cal H}_{k-1}\rightarrow {\cal H}_{k}$, called
the creation operators, are given by
\begin{equation}
\label{a17}
{\bf A}_{k}^{*}=\left ( \partial_{q} +f_{k} \right)^{*}=
B_{k}\left( -\partial_{q}Q^{-1}+f_{k}\right)-A_k\left(1+(1-q)xf_k \right) \;.
\end{equation}
The derivation of the formula (\ref{a17}) is given in Appendix A.
It follows from (\ref{a18}) that the real valued functions $Z_k$, $W_k$ and $V_k$ are 
related to $f_k$, $B_k$, $A_k$ by the formulas:
\begin{align}
\label{a21}
&Z_{k}=-B_{k}Q^{-1} \left( 1+(1-q)id\;f_{k} \right) \; ,\\
\label{a22}
&W_{k}=B_kf_{k}-A_{k}\left( 1+(1-q)id\;f_k \right)-q^{-1}B_{k}Q^{-1}(f_{k}) \;,\\
\label{a23}
&V_{k}=-B_{k}\partial_{q} \left( Q^{-1}(f_{k}) \right) -A_{k}f_{k}\left( 1+(1-q)id\;f_k \right)+B_kf_{k}^{2}+a_k \;\; .
\end{align}
Necessary and sufficient conditions for the consistency of factorization formulas (\ref{a18}) and (\ref{3}) are
\begin{align}
\label{r1}
&\eta_{k+1}(x)=g_k(x) \eta_k(q^{-1}x)\;,\\
\label{r2}
&\varphi_{k+1}(x)=\frac{d_{k+1}}{ g_k(x)} \varphi_k(q^{-1}x)\;,\\
\label{r3}
&\alpha_k(x) -\frac{g_k(qx)}{d_{k+1} }\alpha_k(q x)=
\end{align}$$
=
\left(\frac{q^2 d_{k+1} B_k(qx) -g_k(q^2x) B_k(q^2 x)}{(1-q)^2 q^3 x^2} +d_{k+1}a_k-a_{k+1} \right)
\frac{g_k(qx)}{d^2_{k+1}}\;,
$$
where we have introduced the  additional notations
\begin{align}
\label{r4}
&g_k(x) := \frac{B_{k+1}(x)}{B_k(x)}\;,\\
\label{r5}
&\varphi_k(x) := f_k(x)+ \frac{1}{(1-q) x}\;,\\
\label{r9} &\alpha_k(x):=\varphi_k^2(x) \eta_k(x)\; .
\end{align}
The detailed derivation of these formulas is given in Appendix B and in the paper \cite{1, 38}.

The relations (\ref{r1}), (\ref{r2}) and (\ref{r4}), (\ref{r9}) allow us to express
the functions $B_k$, $\eta_k$,  $\varphi_k$ and $\alpha_k$ by  the initial data
$B_0$, $\eta_0$,  $\varphi_0$ and $\alpha_0$
\begin{align}
\label{r7}
&B_k(x) = g_{k-1}(x) g_{k-2}( x) \dots g_0(x) B_0(x)\;,\\
\label{r6}
&\eta_k(x) = g_{k-1}( x) g_{k-2}(q^{-1} x) \dots g_0(q^{-k+1} x) \eta_0(q^{-k} x)\;,\\
\label{r8}
&\varphi_k(x) = \frac{d_{k} \dots d_1}{g_{k-1}( x) \dots
 g_0(q^{-k+1} x)} \varphi_0(q^{-k} x)\;,\\
\label{r9a1a}
&\alpha_k(x)=\frac{\left(d_{k} \dots d_1\right)^2}{g_{k-1}( x) \dots
 g_0(q^{-k+1} x)}\alpha_0(q^{-k}x)\; .
\end{align}

Substituting (\ref{r7}--\ref{r9a1a}) into condition (\ref{r3})
we obtain the infinite sequence of the nonlinear functional equations
\begin{equation}
\label{r10}
\alpha_0(x) - d_{k+1} \frac{G_{k+1}(x)}{G_k(q x)} \alpha_0(q x) =
G_{k+1}(x)\left( d_{k+1}a_{k}-a_{k+1}+\right.
\end{equation}
$$
+\left.\frac{q^2 d_{k+1} g_{k-1}(q^{k+1} x) \dots g_0(q^{k+1} x) B_0(q^{k+1} x) -
    g_{k}(q^{k+2} x) \dots g_0(q^{k+2} x) B_0(q^{k+2} x)}
{(1-q)^2 q^{2k +3} x^2}\right)\;,
$$
where
\begin{align}
\label{r11}
&G_k(x):= \frac{g_{k-1}(q^k x) \dots g_0(q x)}
{(d_{k} \dots d_1)^2}
\;\;\;\;\;\; \textrm{for }\; k\in {\mathbb N}\;,\\
\label{r12}
&G_0(x) :=1 \;,
\end{align}
for the functions $\alpha_0$, $B_0$ and $g_k$ for $k\in\mathbb{N}\cup \{0\}$.

One sees from (\ref{r7}--\ref{r9a1a}) that the sequence of functions $g_k$,
$k \in {\mathbb N}$, satisfying (\ref{r10}) defines the chain of $q$-- difference operators
(\ref{a15}) if the first element ${\bf H}_0$ of the chain is given. So, the problem of
construction of the  factorized chain given by (\ref{a18}) and (\ref{3}) is equivalent to solving of the system of
functional equations (\ref{r10}).

Let us now present the limit behaviour of the formulas obtained above when the parameter $q$ tends to 1. It is easy to see
that the set $[a,b]_q$ becomes the interval $[a,b]$ in the limit $q\rightarrow 1$ and the scalar product turns to be
\begin{equation}
\langle \psi | \varphi \rangle_k = \int_a^b\overline{\psi(x)}\varphi(x)\varrho_k(x)dx\;,
\end{equation}
where the weight function $\varrho_k(x)$ satisfies the  Pearson equation
\begin{equation}
\frac{d}{dx}(\varrho_kB_k)=\varrho_kA_k\;,
\end{equation}
with  the boundary conditions \eqref{a5}.
For $q\to 1$ the operator $Q$  goes to  the identity operator and
$\partial_q\xrightarrow[q\to 1]{}\frac{d}{dx}$.
In the limiting case the annihilation and creation operators are of the form
\begin{equation} {\bf A}_k=\frac{d}{dx}+f_k,\end{equation}
\begin{equation} {\bf A}_k^*=B_k\left(-\frac{d}{dx}+f_k\right)-A_k\end{equation}
and the operators ${\bf H}_k$ are given by
\begin{equation}
\label{s1x}{\bf H}_k=-B_k \frac{d^2}{dx^2}-A_k\frac{d}{dx}+(f_k^2 -f_k')
B_k-f_kA_k+a_k\; .
\end{equation}
The $q$-difference equation \eqref{a15} tends to the differential equation
\begin{equation}
\left( Z_{k}(x)\frac{d^2}{dx^2}+W_{k}(x)\frac{d}{dx}+V_{k}(x)\right)\psi_k(x)=\lambda_k\psi_k(x) \;,
\end{equation}
where the coefficients are given by
\begin{align}
&Z_{k}(x)=-B_{k}(x)\;,\\
&W_{k}(x)=-A_{k}(x)\;,\\
&V_{k}(x)=\left(  f_{k}^{2}(x)- f'_k(x)\right) B_k(x) -f_k(x)A_k(x)+a_k\; .
\end{align}
The recurrence transformations (\ref{r1}--\ref{r2}) for  $q\to 1$ tend to
\begin{equation}\label{s2x}B_{k+1}=d_{k+1}B_k\;,\end{equation}
\begin{equation}\label{s3x}A_{k+1}=d_{k+1}\left(A_k-\frac{d}{dx} B_k\right).\end{equation}
The sequence of $q$-difference equations \eqref{r3} tends to the sequence of non--linear
differential  equations
\begin{equation}
\label{c6}
B_k(f_{k+1}^2 -f^2_k+f'_{k+1}+f'_k)-A_k(f_{k+1}-{f}_k)+2B'_kf_{k+1} -A'_k+B''_k=
{a}_k-\frac{{a}_{k+1}}{d_{k+1}},
\end{equation}
$k\in\mathbb{N}\cup\{ 0\}$. The equation \eqref{c6} for $B_k(x)\equiv 1$ and $A_k(x)\equiv 0$ was  considered
in many papers (see \cite{2,16,16a,18,7,7b}), but  nevertheless for these differential--difference equations there is no
complete theory.
One of the methods for solving of (\ref{c6}) is to look for  the solutions of (\ref{c6})
in the form of infinite series
\begin{equation}
\label{c6cc}
f_k=\sum_{i\in {\mathbb Z}} \tilde f_i(x)k^i
\end{equation}
and obtain in this way the conditions on the function $\tilde f_i(x)$. The case of
solutions given by the finite series were consider by Infeld and Hull \cite{2}. The classification
of all factorisable one--dimensional problems  is still an open question.

Now, we come back to the general case.
Regarding the extreme nonlinearity of the system (\ref{r10}), the possibility to solve it is
rather out of the question.
Therefore, we shall restrict ourselves to the subcase
\begin{equation}
\label{r18}
g_k(x):= d_{k+1}q^{\gamma}\;\;\;\;\;\; \textrm{ for }\; \gamma\in\mathbb{R}
\end{equation}
and consider the system (\ref{r10}), which is reduced now to
\begin{equation}
\label{r13}
\alpha_0(x)-q^{\gamma} \alpha_0(qx) =
\frac{q^{(k+1)\gamma }}{d_{k+1}\ldots d_1}\left( d_{k+1}a_k-a_{k+_1} \right)+
\end{equation}$$
+q^{2(k+1)\gamma}Q^{k+1}\frac{q^{2-\gamma}  B_0 (x)-B_0(qx)} {(1-q)^2 q x^2} \;,
$$
as the infinite system of equations on the initial functions $B_0$ and $\alpha_0$.
Eliminating $\alpha_0$ from (\ref{r13}) we obtain
\begin{equation}
\label{5}
(1-q)^2q^{3-\gamma}d_1^{-1}x^2\left(\frac{q^{k\gamma}}{d_{k+1}\ldots d_1}
\left(d_{k+1}a_k-a_{k+1}\right)-d_1a_0+a_1 \right)=
\end{equation}$$
=q^{2-\gamma}B_0(qx)-B_0(q^2x)-q^{2k(\gamma -1)}\left( q^{2-\gamma}B_0(q^{k+1}x)-B_0(q^{k+2}x)
\right),\;\;\;\;\;k\in{\mathbb N}\;.
$$
Now, we shall look for the solution of (\ref{5}) in the form
\begin{equation}
\label{6}
B_0(x)=x^\delta\sum_{n\in {\mathbb Z}}b_nx^n\;,
\end{equation}
where $\delta\in \langle 0,1)$. Substituting (\ref{6}) into (\ref{5}) and comparing the
coefficients  in front of $x^n$ we obtain the expressions for the $a_k\in{\mathbb R}$
\begin{equation}
\label{r21abab}
a_{k+1}=d_{k+1}\ldots d_1 q^{-\gamma k}\left(
-a_0 \frac{[\gamma k]}{[\gamma]}+
\frac{a_1}{d_1}\frac{[\gamma (k+1)]}{[\gamma]}-
qb_2[\gamma k][\gamma (k+1)\right), \;\;\;\;\;\;\; k\in {\mathbb N}\;
\end{equation}
and the function $B_0$
\begin{equation}
\label{r26a9}
B_0(x)=b_2x^2+b_1x^{2-\gamma}+b_0x^{2-2\gamma} \;,
\end{equation}
where $b_2,b_1,b_0\in {\mathbb R}$. From (\ref{r26a9}) and (\ref{r13}) we have:
\begin{enumerate}[\bf (i)]
\item if $\gamma \neq 0$, then
\begin{equation}
\label{r30a8}
\alpha_0(x) =
\frac{q^{\gamma+1}b_2}{(1-q)^2}
+  \frac{q^{\gamma}(d_1a_0-a_1)}{(1- q^{\gamma})d_1}+hx^{-\gamma}
+\frac{ q^{1-\gamma}b_0}{(1-q)^2}x^{-2\gamma}\;,
\end{equation}
where $h\in {\mathbb R}$;
\item if $\gamma =0$, then
\begin{equation}
\label{r30a8A}
\alpha_0(x) = h\;\;\; \textrm{and} \;\;\; d_1a_0=a_1\;,
\end{equation}
where $h\in {\mathbb R}$.
\end{enumerate}
Finally, substituting (\ref{r18}) to (\ref{r7}--\ref{r9a1a}) we find the following
transformation formulas
\begin{align}
\label{r17u2}
& B_k(x)= q^{\gamma k}d_{k}\ldots d_1B_0(x) \; ,\\
\label{r17u1}
& \eta_k(x)=q^{\gamma k}d_{k}\ldots d_1\eta_0(q^{-k}x) \;,\\
\label{r17u3}
& \varphi_k(x)= q^{-\gamma k}\varphi_0(q^{-k}x) \;,\\
\label{r9a1a1a}
& \alpha_k(x)=q^{-\gamma k}d_{k} \dots d_1\alpha_0(q^{-k} x)\; ,
\end{align}
where $B_0$, $\alpha_0$  are given by (\ref{r26a9}) and (\ref{r30a8}--\ref{r30a8A}) respectively.
The functions $\eta_0$ and $\varphi_0(x)$ are related to $A_0$ and $\alpha_0$ by
\begin{align}
&\eta_0(x)=b_2x^2 +b_1x^{2-\gamma}+b_0x^{2-2\gamma} -(1-q)xA_0(x)\;,\\
&\varphi_0(x)=\sqrt{\frac{\alpha_0(x)}{\eta_0(x)}}\;.
\end{align}
At the moment, given the functions $B_0$, $\alpha_0$, we can use
(\ref{r17u2}--\ref{r9a1a1a}), (\ref{a6}--\ref{a7}), (\ref{r5}) and (\ref{r9}) in order to
express the functions $A_k$, $f_k$ and $\varrho_k$:
\begin{align}
\label{za1} A_k(x)=& q^{\gamma k}d_{k}\ldots d_1\left(  q^{-k}A_0(q^{-k}x)+[-2k]b_2x+\right.\\
  &\left.+[k(\gamma-2)]b_1x^{1-\gamma}+[2k(\gamma-1)]b_0x^{1-2\gamma}\right),\nonumber\\
\label{za2}
f_k(x)= &q^{-\gamma k}f_0(q^{-k}x)-\frac{1-q^{k(1-\gamma)}}{(1-q)x}\; ,\\
\label{za3} \varrho_k(x)=&  \frac{q^{-\frac{\gamma k(k+1)}{2}}}{d_{k}d^2_{k-1}\ldots d_1^{k}}
\frac{\varrho_0(q^{-k}x)} {\prod_{n=0}^{k-1}\left( b_2q^{-2n}x^2+b_1q^{n(\gamma-2)}x^{2-\gamma}+b_0q^{2n(\gamma-1)}
x^{2-2\gamma} \right) }
\end{align}
by $A_0$, $f_0$ and $\varrho_0$.
From conditions (\ref{a6}--\ref{a7}), (\ref{r5}) and (\ref{r9})
we see that the functions $A_0$, $f_0$, $\varrho_0$ are related by
\begin{align}
\label{r30a10}
&\varrho_0(x)=\frac{q^2b_2x+b_1q^{2-\gamma}x^{1-\gamma}+b_0q^{2(1-\gamma)}x^{1-2\gamma}}
{b_2x+b_1x^{1-\gamma}+b_0x^{1-2\gamma}-(1-q)A_0(x)}\varrho_0(qx)\;, \\
\label{r30b10}
&\left( f_0(x)+\frac{1}{(1-q)x}\right)^2=
\frac{\alpha_0(x)}
{b_2x^2+b_1x^{2-\gamma}+b_0x^{2-2\gamma}-(1-q)xA_0(x)}\;.
\end{align}
So, further we shall assume that the function  $\frac{A_0(0)}{B_0(0)}$ is continuous in $0$. Under this assumption we
obtain from (\ref{r30a10}) and (\ref{r30b10})
\begin{align}
\label{r30a16}
&f_0(x)= \sqrt{\frac{\alpha_0(x) }
{b_2x^2+b_1x^{2-\gamma}+b_0x^{2-2\gamma} -(1-q)xA_0(x)}} -\frac{1}{(1-q)x}\;,\\
\label{r30a17}
&\varrho_0(x)=\frac{1}{ b_2x^2+b_1x^{2-\gamma}+b_0x^{2-2\gamma}}\prod_{n=0}^{\infty}Q^n\left(
\frac{1}{1-(1-q)x\frac{A_0(x)}{b_2x^2+b_1x^{2-\gamma}+b_0x^{2-2\gamma}}}\right)\; .
\end{align}
This means that one finds the explicit  formulas for the annihilation and creation  operators
\begin{align}
\label{r30a18}
{\bf A}_k = & \partial_q-\frac{1}{(1-q)x}+q^{-\gamma k}\sqrt{ \frac{\alpha_0(q^{-k}x) }
{\eta_0(q^{-k}x) }}\;,\\
\label{r30a18a}
{\bf A}_k^*  = & d_k\ldots d_1 \left( -q^{\gamma k}(b_2x^2+b_1x^{2-\gamma}+b_0x^{2-2\gamma})
\left( \partial_qQ^{-1}+\frac{1}{(1-q)x}\right)  +\right.\\
 &\left. +
 \sqrt{ \alpha_0(q^{-k}x)\eta_0(q^{-k}x)}
\right) \nonumber
\end{align}
and from this the explicit expression for ${\bf H}_k$
\begin{equation}
\label{r30a19}
{\bf H}_k=d_k\ldots d_1\left(
- (1-q)q^{-1}x^3(b_2+b_1x^{-\gamma}+b_0x^{-2\gamma})
\sqrt{\frac{\alpha_0(q^{-(k+1)}x)}{\eta_0(q^{-(k+1)}x)} }\partial_q Q^{-1}\partial_q+\right.
\end{equation}
$$
 +\left( - q^{-1}x^2(b_2+b_1x^{-\gamma}+b_0x^{-2\gamma})
\sqrt{\frac{\alpha_0(q^{-(k+1)}x)}{\eta_0(q^{-(k+1)}x)} } + \sqrt{\alpha_0(q^{-k}x)\eta_0(q^{-k}x) }
\right) \partial_q+
$$$$
+\frac{b_2+b_1x^{-\gamma}+b_0x^{-2\gamma}}{(1-q)^2}
\left( q  -  (1-q)x
\sqrt{\frac{\alpha_0(q^{-(k+1)}x)}{\eta_0(q^{-(k+1)}x)} }\right)+
$$$$
+q^{-\gamma k}\alpha_0(q^{-k}x)- \frac{1}{(1-q)x}
\sqrt{\eta_0(q^{-k}x)\alpha_0(q^{-k}x)}+
$$$$
-q^{-\gamma (k-1)}\left( a_0 \frac{[\gamma (k-1)]}{[\gamma]}-
\frac{a_1}{d_1}\frac{[\gamma k]}{[\gamma]}+
qb_2[\gamma (k-1)][\gamma k]\right)
\bigg)\;,
$$
which depend only on a function $A_0$.

The chains of operators ${\bf A}_k$, ${\bf A}_k^*$ and ${\bf H}_k$ appearing in (\ref{r30a18}),
(\ref{r30a18a}) and (\ref{r30a19}) in the limit $q\to 1$ are given by
\begin{equation}{\bf A}_k=\frac d {dx} + f_0(x)+\frac{k(\gamma-1)}{x} \; ,\end{equation}
\begin{equation}{\bf A}_k^*=d_k\ldots d_1\left( B_0(x)\left(-\frac{d}{dx}+f_0(x)+\frac{k(\gamma-1)}{x}\right)-A_0(x)+k\frac{d}{dx}B_0(x)\right)\;,
\end{equation}
\begin{equation}{\bf H}_k=d_k\ldots d_1\left( -B_0(x)\frac{d^2}{dx^2}-(A_0(x)-kB_0'(x))\frac{d}{dx}+\right.
\end{equation}$$
+\left(f_0^2(x)-f_0'(x)+\frac{2k(\gamma-1)}{x}f_0(x)+
\frac{k(\gamma-1)(k(\gamma-1)+1)}{x^2}\right) B_0(x)-
$$$$
\left.-\left(f_0(x)+\frac{k(\gamma-1)}{x}\right)(A_0(x)-kB_0'(x))-
a_0(k-1)+\frac{a_1}{d_1}k-b_2\gamma^2k(k-1)\right) \;,
$$
where
\begin{equation}\label{1y} B_k(x)=d_k\ldots d_1 B_0(x)\; , \end{equation}
\begin{equation}\label{2y} A_k(x)=d_k\ldots d_1 (A_0(x)-k\frac{d}{dx}B_0(x)) \;, \end{equation}
\begin{equation}\label{3y} f_k(x)=f_0(x)+k(\gamma-1)\frac1x \; , \end{equation}
\begin{equation}\label{4y} \varrho_k(x)=\frac{1}{d_kd_{k-1}^2\ldots d_1^k} \frac{\varrho_0(x)}{B_0^k(x)}\;  \end{equation}
and the functions $B_0$, $f_0$ and $\varrho_0$ have the form
\begin{equation}\label{5y} B_0(x)=b_2x^2+b_1x^{2-\gamma}+b_0x^{2-2\gamma}\; , \end{equation}
\begin{equation}\label{6y}
f_0(x)=\left\{
\begin{array}{lc}
\frac{-b_2(\gamma+1)x+\frac{d_1a_0-a_1}{\gamma d_1}x-b_1\tilde{h}
x^{1-\gamma}-b_0(1-\gamma)x^{1-2\gamma}+A_0(x)}{2(b_2x^2+b_1x^{2-\gamma}+b_0x^{2-2\gamma})}
& \textrm{ dla } \gamma \neq 0\\
-\frac{\tilde{\alpha}}{2}\frac{1}{x}+\frac{A_0(x)}{2(b_2+b_1+b_0)x^2} & \textrm{ dla } \gamma=0
\end{array}\right.\;,
\end{equation}
\begin{equation}\label{7y} \varrho_0(x)=\frac{1}{B_0(x)}e^{\int_0^x \frac{A_0(t)}{B_0(t)}dt}\; . \end{equation}

Summing up we see that the construction presented above gives us the nontrivial chain of
 Hamiltonians (\ref{r30a19})  parameterised by the freely chosen function $A_0$
 and the real parameters $b_0$, $b_1$, $b_2$, $\tilde h$, $d_k$ and $\gamma$,
 $k\in {\mathbb N}\cup \{0\}$.

\section{Eigenvalue problem for the chain of operators}

We shall be interested in solving the eigenvalue problems
\begin{equation}
\label{a15d}
{\bf H}_k\psi_k=\lambda_{k}\psi_{k}\;\;\;\textrm{for}\;\;\;k\in{\mathbb N}\cup\{0\}\;.
\end{equation}
If the operators ${\bf H}_k$ admit the factorization given by (\ref{a18}) and (\ref{3}) then the
eigenvalue equation (\ref{a15d}) is equivalent to the two equations
\begin{align}
\label{8}
& {\bf A}_k^*{\bf A}_k\psi_k=(\lambda_k-a_k)\psi_k\;,\\ \label{9}
& {\bf A}_{k+1}{\bf A}_{k+1}^*\psi_k=(d_{k+1}\lambda_k-a_{k+1})\psi_k\;.
\end{align}
From (\ref{8}) and (\ref{9}) one gets
\begin{equation}
\label{10}
{\bf H}_k{\bf A}^*_{k+1}\psi_k=d_{k+1}a_k{\bf A}^*_{k+1}\psi_k
\end{equation}
if
\begin{equation}
\label{10a}
{\bf H}_k\psi_k=a_{k}\psi_{k}\
\end{equation}
or equivalently, if
\begin{equation}
\label{11}
{\bf A}_k\psi_k=0\;.
\end{equation}
Let us remark here that
\begin{equation}
\label{11a}
{\bf A}_{k+1}{\bf A}^*_{k+1}\psi_k=\left(d_{k+1}\lambda_k-a_{k+1}\right)\psi_k
\end{equation}
and thus ${\bf A}^*_{k+1}\psi_k\in{\cal H}_{k+1}$. The formulas (\ref{11a})
show also that the application of  ${\bf A}_{k+1}$ to ${\bf A}^*_{k+1}\psi_k$
turns it  back to the eigenvector of ${\bf H}_k$ proportional to the eigenvector $\psi_k$.
Therefore, in the case when $\lambda_k=a_k$ the eigenvalue problem (\ref{a15d}) is reduced to the equation (\ref{11})
which is a first rank $q$--difference equation, i.e.
\begin{equation}
\label{r17u10u1} \psi_k(x)=\frac{q^{\gamma k} }{(1-q)x   }
\sqrt{\frac{\eta_0(q^{-k}x)}{\alpha_0(q^{-k}x)}}\psi_k(qx),
\end{equation}
where $B_0$ and $\alpha_0$ are given by (\ref{r26a9}) and (\ref{r30a8}--\ref{r30a8A}) respectively.
By applying the iteration method to (\ref{r17u10u1}) we find the solution
\begin{align}
\label{r17u13u1} &\psi_k(x)=x^{\xi_k}  \prod_{n=0}^{\infty}\frac{ q^{\xi_{k}+\gamma k} }
{(1-q)q^nx } \sqrt{\frac{\eta_0(q^{n-k}x)}{\alpha_0(q^{n-k}x)}}\;,
\end{align}
where admissible choices of the real parameter $\xi_k$ and function $A_0$ are presented in the 
table below (Figure \ref{diagram2}). $A(x)$ is to be an arbitrary analytic function.

\begin{figure}[h]
\begin{tabular}{|c|c|c|c|}\hline
 &  & $A_0(x)$ & $\xi_k$ \\
\hline $\gamma > 0$ & \parbox{3.1cm}{$b_0\neq 0$} & $x^{1-2\gamma}A(x)$ &
 $-(\gamma -1)k-\frac{1}{2}\log_q\left( q^{\gamma-1}-(1-q)q^{\gamma-1} \frac{A(0)}{b_0}\right)$  \\ \cline {2-4} &
\parbox{3.1cm}{$b_0=0$ \\ $b_1\neq 0$ \\ $h\neq 0$} & $x^{1-\gamma}A(x)$ &
$-(\gamma -1)k-\frac{1}{2}\log_q\left( \frac{b_1-(1-q)A(0)}{(1-q)^2h}\right)$  \\
\cline {2-4} &\parbox{3.1cm}{$b_0=b_1=h=0$ \\ $b_2\neq 0$ \\ $b_2\neq\frac{(1-q)(a_1-d_1a_0)}{[\gamma]qd_1}$} & $xA(x)$ &
$-(\gamma-1)k-\frac{1}{2}\log_q\left(
\frac{b_2-(1-q)A(0)}{q^{\gamma+1}b_2+\frac{(1-q)q^{\gamma}(d_1a_0-a_1)}{[\gamma]d_1}}\right)$  \\
\hline $\gamma = 0$ &  & $xA(x)$ & $k-\frac{1}{2}\log_q\left(
\frac{b_2+b_1+b_0-(1-q)A(0)}{(1-q)^2\alpha}\right)$\\
 \hline $\gamma< 0$ & \parbox{3.1cm}{ $b_2\neq 0$} & $xA(x)$ &
-($ \gamma -1)k-\frac{1}{2}\log_q\left( \frac{b_2-(1-q)A(0)}{q^{\gamma+1}b_2+\frac{(1-q)q^{\gamma}(d_1a_0-a_1)}
{[\gamma]d_1}}\right)$  \\ \cline{2-4} &
\parbox{3.1cm} {$b_2=0$ \\ $b_1\neq 0$ \\ $h\neq 0$ \\ $d_1a_0=a_1$} & $x^{1-\gamma}A(x)$ &
$-(\gamma -1)k-\frac{1}{2}\log_q\left( \frac{b_1-(1-q)A(0)}{(1-q)^2h}\right)$\\ \cline{2-4} &
\parbox{3.1cm} {$b_2=b_1=h=0$ \\ $b_0\neq 0$ \\ $d_1a_0=a_1$} & $x^{1-2\gamma}A(x)$ &
$-(\gamma -1)k-\frac{1}{2}\log_q\left( q^{\gamma-1}-(1-q)q^{\gamma -1}\frac{A(0)}{b_0}\right)$\\
 \hline
\end{tabular}
\caption{Table of the forms of the function $A_0$ and the parameter $\xi_k$} \label{diagram2}
\end{figure}

Now, let us answer the question of when the solution $\psi_k$ of (\ref{r17u13u1}) belongs to  the
Hilbert space ${\cal H}_k$. In order to do this we observe that
\begin{equation}
\label{A100}
\left( \left|\psi_k\right|^2\varrho_k\right)(x)=\frac{q^{2\gamma k}}{(1-q)^2x^2}\frac{B_0(qx)}
{\alpha_0(q^{-k}x)}\left( \left|\psi_k\right|^2\varrho_k\right)(qx)\;.
\end{equation}
The equation (\ref{A100}) can be written  for $\gamma= 0$ in the form
\begin{equation}
\label{Aa}
\left( \left|\psi_k\right|^2\varrho_k\right)(x)=
\frac{q^2\left(b_2+b_1+b_0\right)}{(1-q)^2\alpha}  \left( \left|\psi_k\right|^2\varrho_k\right)(qx) \;,
\end{equation}
and for $\gamma\neq 0$ in the form
\begin{equation}
\label{Aa1}
\left( \left|\psi_k\right|^2\varrho_k\right)(x)=
\end{equation}
$$
=\frac{ q^{1-\gamma}\left( b_2(qx)^{2\gamma}+b_1(qx)^{\gamma}+b_0\right)}{q^{2\gamma} \left(b_2+
\frac{(1-q)^2}{(1- q^{\gamma})}\frac{(d_1a_0-a_1)}{qd_1} \right)(q^{-k}x)^{2\gamma}
+ (1-q)^2 q^{\gamma -1}h(q^{-k}x)^{\gamma}+b_0 }\left( \left|\psi_k\right|^2\varrho_k\right)(qx)\;.
$$
We also observe that the function $\left|\psi_k\right|^2\varrho_k$ does not depend on $A_0(x)$.
Using iteration method, after standard calculations we obtain the classes of solutions of (\ref{A100})
described in the following proposition.
\begin{lema}
\label{A101}
For the solutions to the equation (\ref{A100}), the following cases hold:\\
1. For $\gamma =0$ we have
\begin{equation}
\left( \left|\psi_k\right|^2\varrho_k\right)(x)        =
x^r\;,
\end{equation}
where $q^{-r}=\frac{q^2\left(b_2+b_1+b_0\right)}{(1-q)^2\alpha}$.\\
2. For $\gamma\neq 0$ we have following possibilities:
\begin{enumerate}[\bf (i)]
\item If $b_0\neq 0$, $b_2\neq 0$ and
$b_2+\frac{(1-q)^2}{(1- q^{\gamma})}  \frac{(d_1a_0-a_1)}{qd_1}    \neq 0$, then
\begin{equation}
\label{A101a}
\left( \left|\psi_k\right|^2\varrho_k\right)(x)        =        x^{\gamma-1}
\frac{\left( \frac{(qx)^{\gamma}}{x_1};q^{\gamma}\right)_{\infty}\left( \frac{(qx)^{\gamma}}{x_2};q^{\gamma}\right)_{\infty}}
{\left( \frac{(q^{-k}x)^{\gamma}}{y_1};q^{\gamma}\right)_{\infty}\left( \frac{(q^{-k}x)^{\gamma}}{y_2};q^{\gamma}\right)_{\infty}}\;.
\end{equation}
\item If $b_0\neq 0$, $b_2\neq 0$, $h\neq 0$ and
$b_2+\frac{(1-q)^2}{(1- q^{\gamma})}   \frac{(d_1a_0-a_1)}{qd_1}    =0$, then
\begin{equation}
\left( \left|\psi_k\right|^2\varrho_k\right)(x)        =           x^{\gamma-1}
\frac{\left( \frac{(qx)^{\gamma}}{x_1};q^{\gamma}\right)_{\infty}\left( \frac{(qx)^{\gamma}}{x_2};q^{\gamma}\right)_{\infty}}
{\left( \frac{(q^{-k}x)^{\gamma}}{y_1};q^{\gamma}\right)_{\infty}}\;.
\end{equation}
\item If $b_0\neq 0$, $b_2\neq 0$, $h=0$ and
$b_2+\frac{(1-q)^2}{(1- q^{\gamma})}   \frac{(d_1a_0-a_1)}{qd_1}    =0$, then
\begin{equation}
\left( \left|\psi_k\right|^2\varrho_k\right)(x)        =   x^{\gamma-1}
\left( \frac{(qx)^{\gamma}}{x_1};q^{\gamma}\right)_{\infty}\left( \frac{(qx)^{\gamma}}{x_2};q^{\gamma}\right)_{\infty}\;.
\end{equation}
\item If $b_0= 0$, $b_1\neq 0$, $b_2\neq 0$, $h\neq 0$ and
$b_2+\frac{(1-q)^2}{(1- q^{\gamma})}    \frac{(d_1a_0-a_1)}{qd_1}    \neq 0$, then
\begin{equation}
\left( \left|\psi_k\right|^2\varrho_k\right)(x)        =  x^r
\frac{\left( \frac{(qx)^{\gamma}}{x_1};q^{\gamma}\right)_{\infty}}
{\left( \frac{(q^{-k}x)^{\gamma}}{y_1};q^{\gamma}\right)_{\infty}}\;,
\end{equation}
where $q^{-r}=\left| \frac{q^{2+\gamma (k-1)}b_1}{(1-q)^2h }\right|$.
\item If $b_0= 0$, $b_1\neq 0$, $b_2\neq 0$, $h\neq 0$ and
$b_2+\frac{(1-q)^2}{(1- q^{\gamma})}     \frac{(d_1a_0-a_1)}{qd_1}    =0$, then
\begin{equation}
\left( \left|\psi_k\right|^2\varrho_k\right)(x)      =
x^{r}\left( \frac{(qx)^{\gamma}}{x_1};q^{\gamma}\right)_{\infty}\;,
\end{equation}
where $q^{-r}=\left|  \frac{q^{2+\gamma (k-1)}b_1}{(1-q)^2h } \right|$.
\item If $b_0=h= 0$, $b_1\neq 0$, $b_2\neq 0$  and
$b_2+\frac{(1-q)^2}{(1- q^{\gamma})}    \frac{(d_1a_0-a_1)}{qd_1}    \neq 0$, then
\begin{enumerate}[(a)]
\item \begin{equation}
\left( \left|\psi_k\right|^2\varrho_k\right)(x)       =
x^{r}\frac{\left( \frac{(qx)^{\gamma}}{x_1};q^{\gamma}\right)_{\infty}}
{\left( -(q^{-k}x)^{\gamma};q^{\gamma}\right)_{\infty} \left( -q^{\gamma}(q^{-k}x)^{-\gamma};q^{\gamma}\right)_{\infty} }\;,
\end{equation}
where $q^{-r}=\frac{q^{k\gamma +1}b_1}
{ q^{2\gamma}\left( b_2+\frac{(1-q)^2}{(1- q^{\gamma})}   \frac{(d_1a_0-a_1)}{qd_1}\right)    }>0$;
\item\begin{equation}
\left( \left|\psi_k\right|^2\varrho_k\right)(x)       =
x^{r}\frac{\left( \frac{(qx)^{\gamma}}{x_1};q^{\gamma}\right)_{\infty}}
{\left( (q^{-k}x)^{\gamma};q^{\gamma}\right)_{\infty} \left( q^{\gamma}(q^{-k}x)^{-\gamma};q^{\gamma}\right)_{\infty} }\;,
\end{equation}
where $-q^{-r}=  \frac{q^{k\gamma +1}b_1}
{ q^{2\gamma}\left( b_2+\frac{(1-q)^2}{(1- q^{\gamma})}   \frac{(d_1a_0-a_1)}{qd_1} \right)   }  <0$.
\end{enumerate}
\item If $b_0=b_1= 0$, $b_2\neq 0$, $h\neq 0$ and
$b_2+\frac{(1-q)^2}{(1- q^{\gamma})}   \frac{(d_1a_0-a_1)}{qd_1}    \neq 0$, then
\begin{enumerate}[(a)]
\item
\begin{equation}
\left( \left|\psi_k\right|^2\varrho_k\right)(x)    =
x^{r}\frac{\left( -x^{\gamma};q^{\gamma}\right)_{\infty} \left( -q^{\gamma}x^{-\gamma};q^{\gamma}\right)_{\infty}}
{\left( \frac{(q^{-k}x)^{\gamma}}{y_1};q^{\gamma}\right)_{\infty} }\;,
\end{equation}
where $q^{-r}= \frac{q^{2+k\gamma}b_2}{(1-q)^2h }>0$;
\item
\begin{equation}
\left( \left|\psi_k\right|^2\varrho_k\right)(x)=
x^{r}\frac{\left( x^{\gamma};q^{\gamma}\right)_{\infty} \left( q^{\gamma}x^{-\gamma};q^{\gamma}\right)_{\infty}}
{\left( \frac{(q^{-k}x)^{\gamma}}{y_1};q^{\gamma}\right)_{\infty}  }\;,
\end{equation}
where $-q^{-r}=\frac{q^{2+k\gamma}b_2}{(1-q)^2h } <0$.
\end{enumerate}
\item If $b_0=b_1=h= 0$, $b_2\neq 0$ and
$b_2+\frac{(1-q)^2}{(1- q^{\gamma})}    \frac{(d_1a_0-a_1)}{qd_1}    \neq 0$, then
\begin{equation}
\left( \left|\psi_k\right|^2\varrho_k\right)(x)=        x^r\;,
\end{equation}
where $q^{-r}=\left| \frac{q^{1-\gamma+2k\gamma}b_2}{b_2+\frac{(1-q)^2}{(1- q^{\gamma})}  \frac{(d_1a_0-a_1)}
{qd_1} }\right|$.
\end{enumerate}
In all the above cases $x_1,x_2$ are roots of the polynomial
\begin{equation}
b_2x^2+b_1x+b_0=0\;
\end{equation}
and $y_1,y_2$ are roots of the polynomial
\begin{equation}
\left( q^{2\gamma}b_2+(1-q)^2\frac{q^{2\gamma-1}(d_1a_0-a_1)}{(1- q^{\gamma})d_1}   \right)x^2
+ (1-q)^2q^{\gamma-1}hx+b_0=0\;.
\end{equation}
\end{lema}
{\bf Proof:} We easily obtain the subcases {\bf (i) -- (iii)} by iteration. The other cases are proved by calculation of the
Laurent expression coefficient and application of Jacobi's identities
\begin{equation}
\sum_{k=-\infty}^{\infty}q^{k^2}x^k=\left( q^2;q^2\right)_{\infty}
\left(-qx;q^2\right)_{\infty}   \left(-q/z;q^2\right)_{\infty}\;,
\end{equation}
(see \cite{34}).\\
\hspace*{15cm}$\square$\\
The proposition given below classifies those function (\ref{r17u13u1}) which are elements of Hilbert space ${\cal H}_k$.
\begin{lema}
\label{A104}
The solution (\ref{r17u13u1}) of equation (\ref{11}) belongs to the Hilbert space
${\cal H}_k$ if and only if the parameters $b_0$, $b_1$, $b_2$, $\alpha$, $h$, $d_1$, $a_0$, $a_1$
and $\gamma$ satisfy the following conditions:
\begin{enumerate}
\item $\gamma=0$ and $\frac{\alpha}{b_2+b_1+b_0}<\frac{q}{(1-q)^2}$.\\
\item $\gamma>0$ and one the following conditions is fulfilled:
\begin{enumerate}[\bf (i)]
\item  $b_0\neq 0$, $b_2\neq 0$ and
$b_2+\frac{(1-q)^2}{(1- q^{\gamma})}  \frac{(d_1a_0-a_1)}{qd_1}    \neq 0$;
\item $b_0\neq 0$, $b_2\neq 0$, $h\neq 0$ and
$b_2+\frac{(1-q)^2}{(1- q^{\gamma})}   \frac{(d_1a_0-a_1)}{qd_1}    =0$;
\item  $b_0\neq 0$, $b_2\neq 0$, $h=0$ and
$b_2+\frac{(1-q)^2}{(1- q^{\gamma})}   \frac{(d_1a_0-a_1)}{qd_1}    =0$;
\item $b_0= 0$, $b_1\neq 0$, $b_2\neq 0$, $h\neq 0$,
$b_2+\frac{(1-q)^2}{(1- q^{\gamma})}    \frac{(d_1a_0-a_1)}{qd_1}    \neq 0$ and
                     $\frac{h}{b_1}<\frac{q^{1+\gamma (k-1)}}{(1-q)^2}$;
\item $b_0= 0$, $b_1\neq 0$, $b_2\neq 0$, $h\neq 0$,
$b_2+\frac{(1-q)^2}{(1- q^{\gamma})}     \frac{(d_1a_0-a_1)}{qd_1}    =0$ and
              $\frac{h}{b_1}<\frac{q^{1+\gamma (k-1)}}{(1-q)^2}$;
\item  $b_0=h= 0$, $b_1\neq 0$, $b_2\neq 0$ and
$b_2+\frac{(1-q)^2}{(1- q^{\gamma})}    \frac{(d_1a_0-a_1)}{qd_1}    \neq 0$;
\item in this case the solutions never belong to the Hilbert space;
\item  $b_0=b_1=h= 0$, $b_2\neq 0$,
$b_2+\frac{(1-q)^2}{(1- q^{\gamma})}    \frac{(d_1a_0-a_1)}{qd_1}    \neq 0$ and
                 $\frac{d_1a_0-a_1}{qd_1b_2}<\frac{1-q^{\gamma}}{(1-q)^2}\left(q^{\gamma (2k-1)}-1)\right)$.
\end{enumerate}
\end{enumerate}
The notation and classification given above are compatible with Proposition \ref{A101}.
\end{lema}
{\bf Proof:}
The function $\psi_k$ belongs to the  Hilbert space if
\begin{equation}
\label{A102}
\int_{[a,b]_q}\left( \left|\psi_k\right|^2\varrho_k\right)(x) d_qx < +\infty \;.
\end{equation}
This is equivalent to
\begin{equation}
\label{A103}
\sum_{n=0}^{\infty}(1-q)q^ny\left( \left|\psi_k^0\right|^2\varrho_k\right)(q^ny)<+\infty
\end{equation}
for $y=a,b$.
So, for the case {\bf (i)} (i.e. $b_0\neq 0$, $b_2\neq 0$ and
$b_2+\frac{(1-q)^2}{(1- q^{\gamma})}  \frac{(d_1a_0-a_1)}{qd_1}    \neq 0$)
we have from Proposition \ref{A101} that the $\left|\psi_k^0\right|^2\varrho_k$ is given by
(\ref{A101a}), and  we  show that
\begin{equation}
\label{Aa2}
(1-q)y^{\gamma}\sum_{n=0}^{\infty}q^{\gamma n}
\frac{\left( \frac{(q^{n+1}y)^{\gamma}}{x_1};q^{\gamma}\right)_{\infty}\left( \frac{(q^{n+1}y)^{\gamma}}{x_2};q^{\gamma}\right)_{\infty}}
{\left( \frac{(q^{n-k}y)^{\gamma}}{y_1};q^{\gamma}\right)_{\infty}\left( \frac{(q^{n-k}y)^{\gamma}}{y_2};q^{\gamma}\right)_{\infty}}<+\infty\;.
\end{equation}
From the identity
\begin{equation}
\left(q^{n\gamma}a;q^{\gamma}\right)_{\infty}=\frac{ \left(a;q^{\gamma}\right)_{\infty}}
{ \left(a;q^{\gamma}\right)_{n}}\;,
\end{equation}
where
\begin{align}
& \left( a;q^{\gamma}\right)_{\infty}=(1-a)(1-q^{\gamma}a)\ldots \;,\\
& \left( a;q^{\gamma}\right)_{n}=(1-a)(1-q^{\gamma}a)\ldots (1-q^{\gamma (n-1)}a)\;,
\end{align}
we obtain the conditions equivalent to (\ref{Aa2})
\begin{equation}
(1-q)y^{\gamma}
\frac{\left( \frac{(qy)^{\gamma}}{x_1};q^{\gamma}\right)_{\infty}\left( \frac{(qy)^{\gamma}}{x_2};q^{\gamma}\right)_{\infty}}
{\left( \frac{(q^{-k}y)^{\gamma}}{y_1};q^{\gamma}\right)_{\infty}\left( \frac{(q^{-k}y)^{\gamma}}{y_2};q^{\gamma}\right)_{\infty}}\times
\end{equation}$$
\times\sum_{n=0}^{\infty}q^{\gamma n}
\frac{\left( \frac{(q^{-k}y)^{\gamma}}{y_1};q^{\gamma}\right)_{n}\left( \frac{(q^{-k}y)^{\gamma}}{y_2}
;q^{\gamma}\right)_{n}}
{\left( \frac{(qy)^{\gamma}}{x_1};q^{\gamma}\right)_{n}\left(
\frac{(qy)^{\gamma}}{x_2};q^{\gamma}\right)_{n}}<+\infty\;.
$$
Those conditions are fulfilled for $\gamma>0$.
The proofs of the other cases are similar to the one above. \\
\hspace*{15cm}$\square$\\

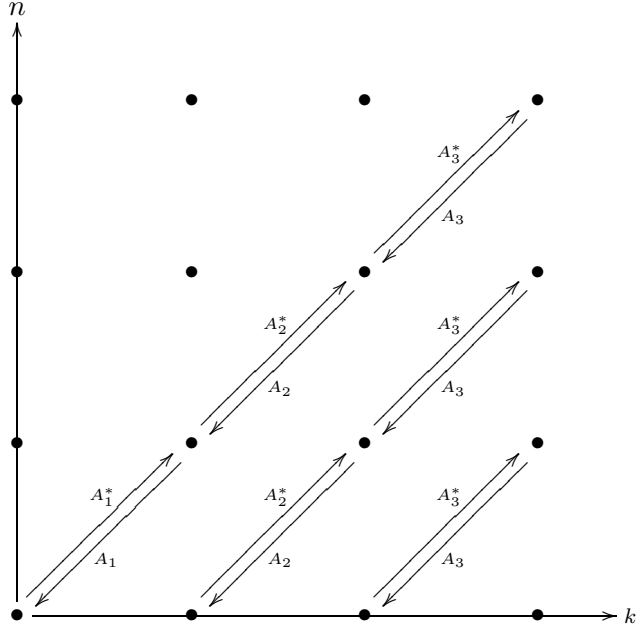
\begin{figure}[h]
\hspace*{4cm}
\xymatrix{
n & & & &  & & & \\
\bullet
& &
\bullet
& &
\bullet
& &
\bullet
\ar@<.5ex>[ddll]^(.6)*-<.9ex>\txt{\tiny{$A_3$}}
& \\
 & & & &  & & & \\
\bullet
& &
\bullet
& &
\bullet
\ar@<.5ex>[uurr]^(.6)*-<.9ex>\txt{\tiny{$A^*_3$}}
\ar@<.5ex>[ddll]^(.6)*-<.9ex>\txt{\tiny{$A_2$}}
& &
\bullet
\ar@<.5ex>[ddll]^(.6)*-<.9ex>\txt{\tiny{$A_3$}}
  & \\
 & & & &  & & & \\
\bullet
& &
\bullet
 \ar@<.5ex>[uurr]^(.6)*-<.9ex>\txt{\tiny{$A^*_2$}}
\ar@<.5ex>[ddll]^(.6)*-<.9ex>\txt{\tiny{$A_1$}}
& &
\bullet
\ar@<.5ex>[uurr]^(.6)*-<.9ex>\txt{\tiny{$A^*_3$}}
\ar@<.5ex>[ddll]^(.6)*-<.9ex>\txt{\tiny{$A_2$}}
& &
\bullet
\ar@<.5ex>[ddll]^(.6)*-<.9ex>\txt{\tiny{$A_3$}}
& \\
 & & & &  & & & \\
\bullet
\ar[rrrrrrr]
\ar@<.5ex>[uurr]^(.6)*-<.9ex>\txt{\tiny{$A^*_1$}}
\ar[uuuuuuu]
& &
\bullet
\ar@<.5ex>[uurr]^(.6)*-<.9ex>\txt{\tiny{$A^*_2$}}
& &
\bullet
\ar@<.5ex>[uurr]^(.6)*-<.9ex>\txt{\tiny{$A^*_3$}}
& &
\bullet
&  _k
}
\caption{Presentation of action of the operators ${\bf A}^*_k$}
\label{diagram1}
\end{figure}

Finally let us come back to the general situation and observe that (\ref{10}), (\ref{10a})
and (\ref{11}) imply that the function
\begin{equation}
\label{r17u11u}
\psi^n_k(x):={\bf A}^*_{k}\ldots {\bf A}^*_{k-n+1}\psi^0_{k-n}(x)\;,\;\;\;\;\;\;n=1, \ldots, k,
\end{equation}
is  an eigenvector of the operator ${\bf H}_k$ with the eigenvalue
\begin{equation}
 \label{r17u12u}
\lambda_k^n=d_kd_{k-1}\ldots d_{k-n+1}a_{k-n}
\end{equation}
if $\psi_{k-n}^0:=\psi_{k-n}$ is the eigenvector of ${\cal H}_{k-n}$ with eigenvalue $a_{k-n}$.
Moreover, one comes back to the eigensubspace ${\mathbb C}\psi_{k-n}^0$ acting on ${\mathbb C}\psi_{k}^n$ by
the annihilation operators
${\bf A}_{k-n+1}$, $\ldots$ and ${\bf A}_{k}$. The above described procedures  can be illustrated
by a lattice of points in the $(k,n)$ plane (Figure \ref{diagram1}).

The eigenfunctions of the operator ${\bf H}_k$ given by (\ref{r17u13u1}) and (\ref{r17u11u}) in the limit  $q\to 1$ tend to
\begin{align} \label{n1} &\psi_k^0(x)=x^{-k(\gamma-1)}e^{-\int_0^x f_0(t)dt} \; , \\
\label{n2}&\psi_k^n(x)={\bf A}_k^*\ldots {\bf A}_{k-n+1}^*x^{-(k-n)(\gamma-1)}e^{-\int_0^x f_0(t)dt}
\;\;\;\textrm{for}\;\;\; n=1,2,\ldots k\; ,
\end{align}
with the eigenvalues
\begin{equation} \lambda_k^n=d_k\ldots d_1 \left(-a_0(k-n-1)+\frac{a_1}{d_1}(k-n)-b_2\gamma^2 (k-n) (k-n-1)\right) \; . \end{equation}

In the next sections we want to present some important examples, including the example of
orthogonal polynomials of $q$--Hahn class which, in the limit $q\rightarrow 1$, gives classical orthogonal polynomials.
These examples will illustrate how the factorization method presented  above works.

\section{ $q$--Hahn orthogonal polynomials  }

We obtain $q$--Hahn orthogonal polynomials when we require that the functions
$f_k(x)\equiv 0$ and $d_k=q^{-1}$ for
$k\in {\mathbb N}\cup\{0\}$.
This is equivalent to
\begin{align}
\label{m2}
&\gamma=1 \;,\\
&B_k(x)=B_0(x)=b_2x^2+b_1x+b_0\; ,\\
&A_0(x)=\left( [2]b_2-qa_0+q^2a_1\right) x+\frac{b_1}{1-q}-(1-q)h\; .
\end{align}
We see that the functions $B_k$ and $A_0$ are  second and a first order polynomials respectively.
From (\ref{za1}) we obtain  that the function $A_k$ is also  first order polynomial
\begin{equation}
\label{m6}
A_k(x)=  q^{-k}A_0(q^{-k}x)+\frac{1-Q^{-k}}{(1-q)x}B_0(x) = \widetilde{a_k} x + \widetilde{b_k}\; ,
\end{equation}
where
\begin{align}
&\widetilde{a_k}=-q^{-2(k-1)}\left( [2(k-1)]b_2+q^{-1}a_0-a_1\right)\;,\\
&\label{ff}\widetilde{b_k}=\frac{b_1}{1-q}-(1-q)q^{-k}h\;.
\end{align}
Hence, the annihilation and creation operators are given by
\begin{align}
\label{m7} & {\bf A}_k=\partial_q\;, \\
\label{m8} & {\bf A}^*_k=- (b_2x^2+b_1x+b_0)\partial_q Q^{-1}-\widetilde{a_k} x - \widetilde{b_k}\;
\end{align}
and the Hamiltonian by
\begin{align}
\label{m9}
H_k = &- (b_2x^2+b_1x+b_0)\partial_q Q^{-1}\partial_q
- \left( \widetilde{a_k} x + \widetilde{b_k}\right) \partial_q   +\\ \nonumber
 & +q^{-2(k-1)}\left(  -q^{-1}a_0[k-1]+a_1[k]-b_2[k-1][k]\right) \; .
\end{align}
The eigenvalue problem for the Hamiltonian (\ref{m9}) is known as the $q$--Hahn equation \cite{37,4}
\begin{equation}
\label{m10}
\left( B_0(x)\partial_qQ^{-1}\partial_q+A_k(x)\partial_q \right)\psi^n_k=\lambda^n_k \psi^n_k\;.
\end{equation}
The eigenvectors 
related to the eigenvalues
\begin{align}
\label{m18} & \lambda^0_k=0\;,\\
\label{m17} & \lambda^n_k=\widetilde{a_k}[n]+b_2[n][n-1]q^{-(n-1)}\; .
\end{align}

are given by
\begin{align}
\label{m11} \psi^0_k&= 1\;,\\
\label{m11a}  \psi^n_k&= {\bf A}_k^*\ldots {\bf A}_{k-n+1}^* 1=
\!\!\prod_{i=k-n+1}^{k}\!\!\!\left( -(b_2x^2+b_1x+b_0)\partial_qQ^{-1}-\widetilde{a}_ix-
\widetilde{b}_i\right) 1\;,
\end{align}
for $k\in {\mathbb N}\cup \{ 0\}$ and $n=1,2,\ldots ,k$.
The functions $\psi_k^0$  (\ref{m11a}) are  polynomials. Each of the families
$\{\psi_k^n\}_{n=0}^k$  is  a system of polynomials orthogonal with respect to the scalar product
given by Jackson's integral
\begin{equation}
\label{m17aa}
\int_{[a,b]_q}\psi^n_k(x)\psi^m_k(x)\varrho_k(x)d_qx \sim \delta_{nm}\;,
\end{equation}
where the weight functions are obtained from (\ref{r30a17})
\begin{equation}
\label{m13}
\varrho_k(x)=\frac{\varrho_0(q^{-k}x)}{B_0(q^{-k+1}x)\ldots B_0(x)}\;.
\end{equation}
The classes of the weight functions $\varrho_0$ and the set of integration $[a,b]_q$ in (\ref{m17aa})
are presented in \cite{4}.

In the limit $q\rightarrow 1$ this case gives us the classical orthogonal polynomials
\begin{equation}
\left(B_0(x)\frac{d^2}{dx^2}+A_k(x)\frac{d}{dx}\right) P^n_k(x)=\lambda^n_k P^n_k(x) \;.
\end{equation}
The functions $B_0$ and $A_k$ are second and first order polynomials given by
\begin{equation} B_k(x)=B_0(x)=b_2x^2+b_1x+b_0 \;, \end{equation}
\begin{equation} A_k(x)=\tilde a_k x + \tilde b_k \;, \end{equation}
where
\begin{equation}\tilde a_k=-2(k-1)b_2 + a_1-a_0\; , \end{equation}
\begin{equation}\tilde b_k=b_1(\tilde h-k)\;,\end{equation}
(in order to obtain this formulas we demand additionally that $h=\frac{b_1q^{\tilde h}}{(1-q)^2}$
in (\ref{ff})).
The eigenvectors $\psi_k^n$ (orthogonal polynomials), in the limiting case, have the forms
\begin{equation}\psi^0_k(x)=1\; ,\end{equation}
\begin{equation}\psi^n_k(x)=\left(B_0(x)\frac{d}{dx}+A_k(x)\right)\left(B_0(x)\frac{d}{dx}+A_{k-1}(x)\right)\ldots\left(B_0(x)\frac{d}{dx}+A_{k-n+1}(x)\right)1\end{equation}
and correspond to the eigenvalues
\begin{equation}\lambda_k^n=\tilde a_k n + b_2n(n-1)\; .\end{equation}

\section{The case of constant weight functions}

We assume that all weight functions are constant $\varrho_k(x)\equiv \operatorname{const}$.
We obtain two cases, which we consider  below
\vspace*{0.5cm}

{\bf $q$--Deformation of the harmonic oscillator }\\
Additionally we demand that $d_k=q^{-1}$ i  $b_0=\varrho_0=1$ for the sake of transparency 
of the formulas.
In this case we have:
\begin{align}
\label{s8}
&\gamma=1\;,\\
&B_k(x)=1\; ,\\
\label{s9}&A_k(x)=0\; ,\\
\label{s10}&f_k(x)= q^{-k}f_0(q^{-k}x)\; ,\\
\label{s11}&\varrho_k=1\; ,
\end{align}
where
\begin{equation}
\label{s12}
f_0(x)=\sqrt{\frac{ q^2(q^{-1}a_0-a_1)}{1- q}+\frac{h}{x}+\frac{1}{(1-q)^2}\frac{1}{x^2} }
-\frac{1}{(1-q)x}\; .
\end{equation}
The annihilation and creation operators are given by
\begin{align}
\label{s13}&{\bf A}_k=\partial_q+q^{-k}f_0(q^{-k}x) \; ,\\
\label{s14}&{\bf A}^*_k= -\partial_qQ^{-1}+ q^{-k}f_0(q^{-k}x) \; .
\end{align}
Solving the equation (\ref{r17u10u1})
we find the basic state $\psi_k^0$ of the Hamiltonian given by
\begin{equation}
\label{s20} {\bf H}_k=-\left( 1 +(1-q)q^{-k-1}xf_0(q^{-k-1}x) \right) \partial_qQ^{-1}\partial_q+
\end{equation}
$$+
q^{-k} \left( f_0(q^{-k}x)-q^{-1}f_0(q^{-k-1}x) \right) \partial_q+
$$
$$
 -q^{-k}\partial_q(f_0(q^{-k-1}x))+q^{-2k}f_0^2(q^{-k}x) +
q^{-2k}\left(a_0+(q^2a_1-a_0)[k] \right)\;.
$$
\begin{enumerate}
\item If $a_0\neq qa_1$, then
\begin{equation}
\label{s15}
\psi_k^0(x)=\frac{C^0_k}{\sqrt{\left( \frac{q^{-k}x}{x_1}; q\right)_{\infty}
\left( \frac{q^{-k}x}{x_2}; q\right)_{\infty}   }}\;,
\end{equation}
where  $x_1$ and $x_2$ are roots of the polynomial
\begin{equation}
\label{s16}
(1-q)q^2(q^{-1}a_0-a_1)x^2+(1-q)^2hx+1=0
\end{equation}
and $C^0_k\in{\mathbb R}\setminus \{ 0\}$.
\item If $a_0=qa_1$ i $h\neq 0$, then
\begin{equation}
\label{s17}
\psi_k^0(x)=\frac{C^0_k}{\sqrt{\left( -(1-q)^2hq^{-k}x; q\right)_{\infty}   }}\;.
\end{equation}
\end{enumerate}
It easy to see that the operator  $Q^{-1}$ acts as follows

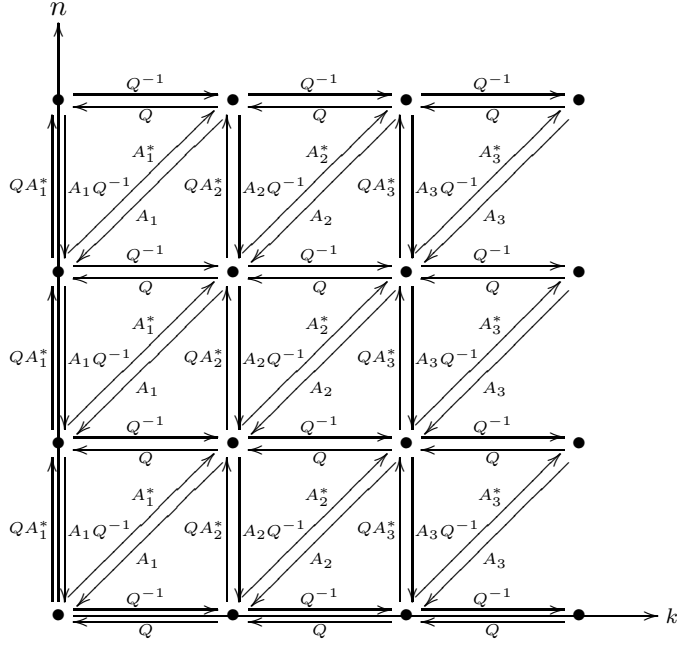
\begin{figure}[ht]
\hspace*{2cm}
\xymatrix{
n & & & &  & & & \\
\bullet
\ar@<.5ex>[dd]^*-<.9ex>\txt{\tiny{$A_1Q^{-1}$}}
\ar@<.5ex>[rr]^*-<0.9ex>\txt{\tiny{$Q^{-1}$}}
& &
\bullet
\ar@<.5ex>[dd]^*-<.9ex>\txt{\tiny{$A_2Q^{-1}$}}
\ar@<.5ex>[rr]^*-<0.9ex>\txt{\tiny{$Q^{-1}$}}
\ar@<.5ex>[ddll]^(.6)*-<.9ex>\txt{\tiny{$A_1$}}
\ar@<.5ex>[ll]^*-<0.9ex>\txt{\tiny{$Q$}}
& &
\bullet
\ar@<.5ex>[dd]^*-<.9ex>\txt{\tiny{$A_3Q^{-1}$}}
\ar@<.5ex>[rr]^*-<0.9ex>\txt{\tiny{$Q^{-1}$}}
\ar@<.5ex>[ddll]^(.6)*-<.9ex>\txt{\tiny{$A_2$}}
\ar@<.5ex>[ll]^*-<0.9ex>\txt{\tiny{$Q$}}
& &
\bullet
\ar@<.5ex>[ddll]^(.6)*-<.9ex>\txt{\tiny{$A_3$}}
\ar@<.5ex>[ll]^*-<0.9ex>\txt{\tiny{$Q$}}
& \\
 & & & &  & & & \\
\bullet
\ar@<.5ex>[dd]^*-<.9ex>\txt{\tiny{$A_1Q^{-1}$}}
\ar@<.5ex>[uu]^*-<.9ex>\txt{ \tiny{$QA^*_1$} }
\ar@<.5ex>[uurr]^(.6)*-<.9ex>\txt{\tiny{$A^*_1$}}
\ar@<.5ex>[rr]^*-<0.9ex>\txt{\tiny{$Q^{-1}$}}
& &
\bullet
\ar@<.5ex>[uu]^*-<.9ex>\txt{ \tiny{$QA^*_2$} }
\ar@<.5ex>[uurr]^(.6)*-<.9ex>\txt{\tiny{$A^*_2$}}
\ar@<.5ex>[ddll]^(.6)*-<.9ex>\txt{\tiny{$A_1$}}
\ar@<.5ex>[rr]^*-<0.9ex>\txt{\tiny{$Q^{-1}$}}
\ar@<.5ex>[ll]^*-<0.9ex>\txt{\tiny{$Q$}}
\ar@<.5ex>[dd]^*-<.9ex>\txt{\tiny{$A_2Q^{-1}$}}
& &
\bullet
\ar@<.5ex>[uu]^*-<.9ex>\txt{ \tiny{$QA^*_3$} }
\ar@<.5ex>[uurr]^(.6)*-<.9ex>\txt{\tiny{$A^*_3$}}
\ar@<.5ex>[ddll]^(.6)*-<.9ex>\txt{\tiny{$A_2$}}
\ar@<.5ex>[rr]^*-<0.9ex>\txt{\tiny{$Q^{-1}$}}
\ar@<.5ex>[ll]^*-<0.9ex>\txt{\tiny{$Q$}}
\ar@<.5ex>[dd]^*-<.9ex>\txt{\tiny{$A_3Q^{-1}$}}
& &
\bullet
\ar@<.5ex>[ddll]^(.6)*-<.9ex>\txt{\tiny{$A_3$}}
\ar@<.5ex>[ll]^*-<0.9ex>\txt{\tiny{$Q$}}
  & \\
 & & & &  & & & \\
\bullet
\ar@<.5ex>[dd]^*-<.9ex>\txt{\tiny{$A_1Q^{-1}$}}
\ar@<.5ex>[uu]^*-<.9ex>\txt{ \tiny{$QA^*_1$} }
\ar@<.5ex>[uurr]^(.6)*-<.9ex>\txt{\tiny{$A^*_1$}}
\ar@<.5ex>[rr]^*-<0.9ex>\txt{\tiny{$Q^{-1}$}}
& &
\bullet
\ar@<.5ex>[uu]^*-<.9ex>\txt{ \tiny{$QA^*_2$} }
\ar@<.5ex>[uurr]^(.6)*-<.9ex>\txt{\tiny{$A^*_2$}}
\ar@<.5ex>[ddll]^(.6)*-<.9ex>\txt{\tiny{$A_1$}}
\ar@<.5ex>[rr]^*-<0.9ex>\txt{\tiny{$Q^{-1}$}}
\ar@<.5ex>[ll]^*-<0.9ex>\txt{\tiny{$Q$}}
\ar@<.5ex>[dd]^*-<.9ex>\txt{\tiny{$A_2Q^{-1}$}}
& &
\bullet
\ar@<.5ex>[uu]^*-<.9ex>\txt{ \tiny{$QA^*_3$} }
\ar@<.5ex>[uurr]^(.6)*-<.9ex>\txt{\tiny{$A^*_3$}}
\ar@<.5ex>[ddll]^(.6)*-<.9ex>\txt{\tiny{$A_2$}}
\ar@<.5ex>[rr]^*-<0.9ex>\txt{\tiny{$Q^{-1}$}}
\ar@<.5ex>[ll]^*-<0.9ex>\txt{\tiny{$Q$}}
\ar@<.5ex>[dd]^*-<.9ex>\txt{\tiny{$A_3Q^{-1}$}}
& &
\bullet
\ar@<.5ex>[ddll]^(.6)*-<.9ex>\txt{\tiny{$A_3$}}
\ar@<.5ex>[ll]^*-<0.9ex>\txt{\tiny{$Q$}}
& \\
 & & & &  & & & \\
\bullet
\ar[rrrrrrr]
\ar@<.5ex>[uu]^*-<.9ex>\txt{ \tiny{$QA^*_1$} }
\ar@<.5ex>[uurr]^(.6)*-<.9ex>\txt{\tiny{$A^*_1$}}
\ar@<.5ex>[rr]^*-<0.9ex>\txt{\tiny{$Q^{-1}$}}
\ar[uuuuuuu]
& &
\bullet
\ar@<.5ex>[uu]^*-<.9ex>\txt{ \tiny{$QA^*_2$} }
\ar@<.5ex>[uurr]^(.6)*-<.9ex>\txt{\tiny{$A^*_2$}}
\ar@<.5ex>[rr]^*-<0.9ex>\txt{\tiny{$Q^{-1}$}}
\ar@<.5ex>[ll]^*-<0.9ex>\txt{\tiny{$Q$}}
& &
\bullet
\ar@<.5ex>[uu]^*-<.9ex>\txt{ \tiny{$QA^*_3$} }
\ar@<.5ex>[uurr]^(.6)*-<.9ex>\txt{\tiny{$A^*_3$}}
\ar@<.5ex>[rr]^*-<0.9ex>\txt{\tiny{$Q^{-1}$}}
\ar@<.5ex>[ll]^*-<0.9ex>\txt{\tiny{$Q$}}
& &
\bullet
\ar@<.5ex>[ll]^*-<0.9ex>\txt{\tiny{$Q$}}
&  _k
}
\caption{Presentation of action of the operators}
\label{diagram4}
\end{figure}

\hspace*{3cm}
\xymatrix{
\psi^{0}_{0} \ar[r]^{\frac{C_1^0}{C_0^0}Q^{-1}} &  \psi^{0}_{1}
\ar[r]^{\frac{C_2^0}{C_1^0}Q^{-1}} &  \ldots  \ar[r]^{\frac{C_k^0}{C_{k-1}^0}Q^{-1}} &
\psi^{0}_{k}  \ar[r]^{\frac{C_{k+1}^0}{C_k^0}Q^{-1}} & \ldots \;, &
}
\\
or equivalently
\begin{equation}
\label{s19}
\psi_k^0(x)=Q^{-k}\psi^0_0(x)\; .
\end{equation}
The functions $\psi_k^0$ are eigenvectors of the Hamiltonians ${\bf H}_k$ with the eigenvalues
\begin{equation}
\label{s21}
\lambda_k^0=a_k=q^{-2k}\left(a_0+(q^2a_1-a_0)[k]\right)\;.
\end{equation}
Similarly it is easy to obtain that the functions
\begin{equation}
\label{s22}
\psi_k^n(x)=Q^{-k}\psi_0^n(x)
\end{equation}
are eigenvectors of ${\bf H}_k$ with
\begin{equation}
\label{s23}
\lambda_k^n=  q^{-2k}\left( \lambda^n_0 +(q^2a_1-a_0)[k]\right)\;,
\end{equation}
in view the following  commutation relations
\begin{align}
\label{s27} & q{\bf A}^*_kQ^{-1}=Q^{-1}{\bf A}^*_{k-1}\; ,\\
\label{s28} & {\bf A}^*_kQ=qQ{\bf A}^*_{k+1}\; ,\\
\label{s29} & q{\bf A}_kQ^{-1}=Q^{-1}{\bf A}_{k-1}\; ,\\
\label{s30} & {\bf A}_kQ=qQ{\bf A}_{k+1}\; .
\end{align}
Finally we present the action of the operators diagrammatically Figure \ref{diagram4} and
state the following 

\begin{lema}
\label{3l}
The functions
\begin{equation}
\label{s41}
\psi_k^n(x)=\frac{1}{\sqrt{(a_0-qa_1)^nn_q!q^{n(n-1)+k}}}Q^{n-k}{\bf A}^*_n\ldots {\bf A}^*_{1}\psi_0^0(x)\;,
\end{equation}
for $k\in{\mathbb N}\cup \{ 0 \}$ and $n\in{\mathbb N}\cup \{ 0 \}$, where the function $\psi_0^0$
is given by (\ref{s15}) or (\ref{s17}), are the eigenvectors of  Hamiltonians (\ref{s20})
corresponding to the eigenvalues
\begin{equation}
\label{s42}
\lambda_k^n=q^{-2k+n}\left(a_0+(q^2a_1-a_0)[k-n]\right)\;.
\end{equation}
\end{lema}

In the limit $q\rightarrow 1$ this case gives us the harmonic oscillator
\begin{equation} {\bf H}_k=-\frac{d^2}{dx^2}+\frac{(a_0-a_1)^2}{4}x^2+\frac{a_1+a_0}{2}+(a_1-a_0)k\;.\end{equation}
with eigenvectors
\begin{equation} \psi_k^n(x)=\left( -\frac{d}{dx}+\frac{a_0-a_1}{2}x\right)^n  e^{-\frac{a_0-a_1}{4}x^2}
\;\;\;\textrm{for}\;\;\;n\in{\mathbb N}\cup\{0\}
\end{equation}
corresponding to the eigenvalues
\begin{equation} \lambda_k^n=a_0+(a_0-a_1)(n-k)\;.\end{equation}
\vspace*{0.5cm}

{\bf $q$--Deformation of the three--dimensional isotropic harmonic oscillator}\\
Additionally we demand that $d_k=q^{-2}$ and $b_1=\varrho_0=1$.
In this case we have
\begin{align}
\label{s43}
&\gamma =2\;,\\
&B_k(x)=1\; ,\\
\label{s44}&A_k(x)=0\; ,\\
\label{s45}&f_k(x)=q^{-2k}f_0(q^{-k}x)-\frac{1-q^{-k}}{(1-q)x}\; ,\\
\label{s46}&\varrho_k=1\; ,
\end{align}
where
\begin{equation}
\label{s47}
f_0(x)=\sqrt{\frac{ q^4(q^{-2}a_0-a_1)}{1- q^2}+\frac{h}{x^2}}
 -\frac{1}{(1-q)x}\; .
\end{equation}
The annihilation and creation operators have the form
\begin{align}
\label{s48}&{\bf A}_k=\partial_q+q^{-2k}f_0(q^{-k}x) -\frac{1- q^{-k}}{(1-q)x} \; ,\\
\label{s49}&{\bf A}^*_k= -\partial_qQ^{-1}+q^{-2k}f_0(q^{-k}x)-\frac{1-q^{-k}}{(1-q)x}   \;,
\end{align}
and the Hamiltonians are given by the formulas
\begin{equation}
\label{s54}
{\bf H}_k=- \left( q^{-k}+(1-q)q^{-2k-1}xf_0(q^{-k-1}x)\right) \partial_qQ^{-1}\partial_q+
\end{equation}
$$ +q^{-2k}\left(f_0(q^{-k}x)-q^{-1}f_0(q^{-k-1}x)\right) \partial_q
-q^{-2k}\left(\partial_qf_0(q^{-k-1}x)\right)+\frac{q^{-2k}[k][k+1]}{x^2}+
$$$$
+q^{-4k}f_0^2(q^{-k}x)+2q^{-3k}\frac{[k]}{x}f_0(q^{-k}x)+
q^{-4k}\left( a_0+(q^4a_1-a_0)\frac{[2k]}{[2]}\right)\;,
$$
The basic states of the Hamiltonians (\ref{s54}) can be found as the solution (\ref{r17u10u1}).
\begin{enumerate}
\item If $a_0\neq q^2a_1$, then
\begin{equation}
\label{s50}
\psi_k^0(x)=\frac{C^0_k}{\sqrt{\left( -\frac{q^4(q^{-2}a_0-a_1)}{(1-q^2)h}q^{-2k}x^2;q\right)_{\infty}}}x^{\xi_k}\;,
\end{equation}
where $C^0_k\in {\mathbb R}\setminus \{ 0\}$ and
\begin{equation}
\label{s51}
\xi_k=-k+\log_q (1-q)\sqrt{h}\;.
\end{equation}
\item If $a_0=q^2a_1$, then
\begin{equation}
\label{s52}
\psi_k^n(x)=C^0_kx^{\xi_k}\;.
\end{equation}
\end{enumerate}
These are the eigenfunctions of the Hamiltonian corresponding to the eigenvalues
\begin{equation}
\label{s55}
\lambda_k^0=a_k=q^{-4k}\left( a_0+(q^4a_1-a_0)\frac{[2k]}{[2]}\right) \;.
\end{equation}
Finally we have the following lemma:
\begin{lema}
\label{6l}
The functions
\begin{equation}
\label{s65}
\psi_k^n(x)={\bf A}^*_k\ldots {\bf A}^*_{k-n+1}\psi^0_{k-n}=
\end{equation}
$$
=\prod_{i=k-n+1}^k\left(  \frac{1}{(1-q)x}\left( -Q^{-1}+
q^{-k}(1-q)\sqrt{h}
\sqrt{1+\frac{q^4(q^{-2}a_0-a_1)}{(1-q^2)h}q^{-2k}x^2}
\right) \right)\psi_{k-n}^0\;,
$$
for $n=1,2,\ldots ,k$,  are the eigenvectors of  the Hamiltonian with the eigenvalues
\begin{equation}
\label{s66}
\lambda_k^n=q^{-2n}a_{k-n}=q^{-2(2k-n)}\left(a_0+(q^4a_1-a_0)\frac{[2(k-n)]}{[2]}\right)\;.
\end{equation}
\end{lema}

In the limit $q\rightarrow 1$ this case gives us the three--dimensional isotropic harmonic oscillator
\begin{equation} {\bf H}_k=-\frac{d^2}{dx^2}+\frac{(k-\frac{\tilde h}{2})(k-\frac{\tilde h}{2}+1)}{x^2}+
\frac{(a_0-a_1)^2}{16}x^2-
\end{equation}$$-
\frac{a_0-a_1}{2}(k+\frac{\tilde h}{2})+\frac{3a_0+a_1}{4}\;.$$
with eigenvectors
\begin{align}
&\psi^0_k(x)=C^0_k x^{\frac{\tilde h}{2}-k}e^{-\frac{a_0-a_1}{8}x^2}\;,\\
&\psi_k^n(x)=\prod^{k}_{i=k-n+1}\left( -\frac{d}{dx}+\frac{a_0-a_1}{4}x-\frac{\tilde h}{2}\frac1x+\frac ix
\right)  x^{\frac{\tilde h}{2}-k}e^{-\frac{a_0-a_1}{8}x^2}
\;\;\;\textrm{for}\;\;\;n=1,\ldots, k
\end{align}
corresponding to the eigenvalues
\begin{equation} \lambda_k^n=a_0+(a_1-a_0)(k-n)\;.\end{equation}

\section*{Appendix A. Derivation of  formula (\ref{a17})}

By the definition of the adjoint operator we have
$$
\left\langle Q^*\psi_k|\varphi_k\right\rangle_k=\left\langle\psi_k|Q\varphi_k\right\rangle_k=
\int_a^b\overline{\psi_k(x)}\varphi_k(qx)\varrho_k(x)d_qx=
$$$$
=\sum_{n=0}^{\infty}(1-q)q^nb\overline{\psi_k(q^{n}b)}\varphi_k(q^{n+1}b)\varrho_k(q^nb)-
\sum_{n=0}^{\infty}(1-q)q^na \overline{\psi_k(q^{n}a)}\varphi_k(q^{n+1}a)\varrho_k(q^na)=
$$$$
\stackrel{m=n+1}{=}
\sum_{m=1}^{\infty}(1-q)q^{m}bq^{-1}\overline{\psi_k(q^{m-1}b)}\varphi_k(q^{m}b)\varrho_k(q^{m-1}b)-
$$$$-
\sum_{m=1}^{\infty}(1-q)q^maq^{-1} \overline{\psi_k(q^{m-1}a)}\varphi_k(q^{m}a)\varrho_k(q^{m-1}a)\;.
$$
In this sum the expression for $m=0$, i.e.
\begin{equation}
 (1-q)\left( b\overline{\psi_k(q^{-1}b)}\varphi_k(b)\varrho_k(b)-
 a\overline{\psi_k(q^{-1}a)}\varphi_k(a)\varrho_k(a)\right)\;,
\end{equation}
does not appear.
The functions $\psi_k(x)$ i $\varphi_k(x)$  are defined on the set
$\{ q^nb:n\in {\mathbb N}\cup\{0\}\}\cup \{ q^na:n\in {\mathbb N}\cup\{0\}\} $ and for the other points
we shall put these functions equal to zero
 $$\begin{array}{cc}
 \left( Q^{-1}\psi\right)(b)=0 \; , \\
 \left( Q^{-1}\psi\right)(a)=0 \;.
 \end{array}
 $$
From the equation (\ref{a4}) we obtain for $x\neq a$ and $x\neq b$ that
\begin{equation}
\left\langle Q^*\psi_k|\varphi_k\right\rangle_k=
\int_{a}^{b}\overline{\psi_k(q^{-1}x)}\varphi_k(x)\varrho_k(x)\frac{B_k(x)}{\eta_k(q^{-1}x)}
q^{-1}d_qx=\left\langle q^{-1}\frac{B_k}{Q^{-1}\eta_k}
(Q^{-1}\psi_k)\bigg|\varphi_k\right\rangle_k \;.
\end{equation}

Similarly we have
$$
\langle f\psi_k|\varphi_{k-1}\rangle_{k-1}=
\int_a^b\overline{f(x)}\overline{\psi_k(x)}\varphi_{k-1}(x)\varrho_{k-1}(x)d_qx=
$$
\begin{equation}
=\int_a^b\overline{\psi_k(x)}\overline{f(x)}\varphi_{k-1}(x)\eta_k(x)\varrho_{k}(x)d_qx=
\langle\psi_k|\overline{f}\eta_k\varphi_{k-1}\rangle_{k} \;,
\end{equation}
where we use the equation (\ref{a2}).

Summarising we obtain the formula (\ref{a17})
\begin{equation}
\label{a17aqa}
{\bf A}_{k}^{*}=\left ( \partial_{q} +f_{k} \right)^{*}=
B_{k}\left( -\partial_{q}Q^{-1}+f_{k}\right)-A_k\left(1+(1-q)xf_k \right) \;,
\end{equation}
where the operator $Q^{-1}$ is given by
\begin{equation}
\label{R5}
Q^{-1}\varphi (x)=
\left\{
\begin{array}{ccll}
\varphi (q^{-1}x) & \textrm{ dla } & x\neq a \textrm{ i } x\neq  b\\
0 & \textrm{ dla } & x=a \textrm { lub } x=b
\end{array}
\right. \; .
\end{equation}

\section*{Appendix B. Derivation of  formulas (\ref{r1}--\ref{r3})}

The operators of annihilation and creation given by (\ref{a16}, \ref{a17}) can be rewritten in the form
\begin{align}
\label{A10} & {\bf A}_{k}=\partial_{q} +f_{k}=-\frac{1}{(1-q)x}Q+\varphi_k\;,\\
\label{A11} & {\bf A}_{k}^{*}=B_{k}\left( -\partial_{q}Q^{-1}+f_{k}\right)-A_k\left(1+(1-q)xf_k \right)=
-\frac{B_k}{(1-q)x}Q^{-1}+\eta_k\varphi_k \;,
\end{align}
where the functions $\varphi_k $, $\eta_k$ are defined by (\ref{r5}) and (\ref{a6}).
From the conditions (\ref{a18}--\ref{3}) we have that
\begin{equation}
\label{A12}
\left\{
\begin{array}{l}
\eta_k(qx)\varphi_k(qx)=d_k\eta_{k-1}(x)\varphi_{k-1}(x)\\
B_k(x)\varphi_k(x)=d_kB_{k-1}(x)\varphi_{k-1}(q^{-1}x)\\
\eta_k(x)\varphi_k^2(x)-d_k\eta_{k-1}(x)\varphi^2_{k-1}(x)=d_ka_{k-1}-a_k+
\frac{q^2d_kB_{k-1}(x)-B_k(qx)}{(1-q)^2qx^2}\;.
\end{array}\right.
\end{equation}
The first and second equations  of  (\ref{A12}) are equivalent to
\begin{align}
& \frac{\varphi_k(qx)}{\varphi_{k-1}(x)}=d_k\frac{\eta_{k-1}(x)}{\eta_k(qx)}\;,\\
& \frac{\varphi_k(qx)}{\varphi_{k-1}(x)}=d_k\frac{B_{k-1}(qx)}{B_k(qx)}\;.
\end{align}
A simple calculation gives us
\begin{align}
\label{A13} & \eta_k(x)=\frac{B_k(x)}{B_{k-1}(x)}\eta_{k-1}(q^{-1}x)=g_{k-1}(x)\eta_{k-1}(q^{-1}x)\;,\\
\label{A14} & \varphi_k(x)=d_k\frac{B_{k-1}(x)}{B_{k}(x)}\varphi_{k-1}(q^{-1}x)=\frac{d_k}{g_{k-1}(x)}\varphi_{k-1}(q^{-1}x)\;,
\end{align}
where the function $g_k(x)$  is given by (\ref{r4}).
Substituting  (\ref{A13}, \ref{A14}) into the third relation in (\ref{A12}) we obtain finally
\begin{equation}
\eta_{k-1}(x)\varphi_{k-1}^2(x)-\frac{g_{k-1}(qx)}{d_k}\eta_{k-1}(qx)\varphi^2_{k-1}(qx)=
\end{equation}
$$
\left( d_ka_{k-1}-a_k+\frac{q^2d_kB_{k-1}(qx)-g_{k-1}(q^2x)B_{k-1}(q^2x)}{(1-q)^2q^3x^2}\right) \frac{g_{k-1}(qx)}{d^2_{k}}\;.
$$

\section*{Acknowledgement}
The authors would like to thank T. Goliński for careful reading of the manuscript and
interest in the paper.

\bibliographystyle{plain}

\begin{thebibliography}{aaaaaa}
\bibitem{40} Chihara, T. S., {\it An Introduction to Orthogonal Polynomials},
             New York: Gordon and Breach, 1978;
\bibitem{9}   Darboux, G., {\it Sur une proposition relative aux equations lineaires}, C. R. Acad.
            Sci. Paris, 94 1456, 1882;
\bibitem{11}  Dirac, P.A.M., {\it Principles of Quantum Mechanics}, Clarendon Press, Oxford, 1947;
\bibitem{34} Gasper, G. and Rahman, M., {\it Basic Hypergeometric Series},
            Cambridge: Cambridge University Press, 1990;
\bibitem{38} Goliński, T., Odzijewicz, A., {\it General difference calculus and its
             application to functional equations of the second order}, Czechoslovak
         J. of Phys., {\bf 52}, 1219-1224, 2002;
\bibitem{37} Hahn, W., {\it \"Uber Orthogonalpolynome die q-Differenzengleichungeg gen\"ungen},
             Math. Nachr., {\bf 2}, 4-34, 1949;
\bibitem{2} Infeld, L. and Hull, T.E., {\it The Factorization Method},
            Rev.  Mod. Phys., {\bf 23}, 21-68, 1951;
\bibitem{47} Koekoek, R. and  Swarttouw, R.F., {\it The Askey--scheme of hypergeometric orthogonal
            polynomials and its $q$--analogue}, Report no. 98-17, TUDelft, webpage
        http://aw.twi.tudelft.nl/koekoek/askey.html, 1998;
\bibitem{16}   de Lange, O.L. and  Raab, R.E., {Operator Methods in Quantum Mechanics}, Claredon Press -
            Oxford, 1991;
\bibitem{16a} Mielnik, B., Nieto, L.M., Rosas-Ortiz, O., {\it The finite difference
             algorithm for higer order supersymmetry}, Phys. Lett. A, {\bf 269}, 70-78, 2000;
\bibitem{18} Miller, W., Jr., {\it Lie Theory and Special Functions}, Academic Press  New York
            and London, 1968;
\bibitem{1} Odzijewicz, A.,  Goliński, T., {\it Second order functional equations};
             webpage  http://arXiv:math-ph/0208006, 2002;
\bibitem{4} Odzijewicz, A., Horowski, H., Tereszkiewicz, A., {\it Integrable
            multi-boson systems and orthogonal polynomials},  J. Phys. A: Math. Gen.,
        {\bf 34}, 4353-4376, (2001);
\bibitem{8}  Schr\"odinger, E., Proc. Roy Irish Acad., A {\bf 46}, 1940;
\bibitem{7}  Spiridonov, V., {\it Universal superpositions of coherent states and self--similar
            potentials}, Phys. Rev. A, {\bf 52}, 1909-1935, 1995;
\bibitem{7b} Veselov, A.P., Shabat, A.B., Funct. Anal. Appl., {\bf 27}, 1-21, 1993;
\end{thebibliography}

\end{document}